\documentclass[twocolumn,showpacs,prb,amsfonts,amsmath,amssymb,floatfix]{revtex4} 
\usepackage{color}
\usepackage{hhline}
\usepackage{mathrsfs}
\usepackage{graphicx}
\usepackage{dcolumn}
\usepackage{bm}
\usepackage{multirow}
\usepackage{booktabs}
\usepackage{afterpage}
\usepackage{amsmath}
\usepackage[T1]{fontenc}

\begin{document}

\title{Electron-hole dichotomy and enhancement of thermoelectric power factor by electron-hole-asymmetric relaxation time: a model study on a two-valley system with strong intervalley scattering}

\author{Masayuki Ochi}
\affiliation{Forefront Research Center, Osaka University, Machikaneyama-cho, Toyonaka, Osaka 560-0043, Japan}
\affiliation{Department of Physics, Osaka University, Machikaneyama-cho, Toyonaka, Osaka 560-0043, Japan}

\date{\today}
\begin{abstract}
The role of electron-phonon scattering in thermoelectric transport has been paid much attention, especially in multivalley systems.
By investigating a two-valley model with electron-phonon coupling,
we find three electron transport regimes realized by electron-hole asymmetry of electron relaxation time due to the strong intervalley scattering.
Seebeck coefficient denotes an electron-hole dichotomy due to this asymmetry.
Also, the strong intervalley scattering can enhance power factor.
Our finding sheds light on unexplored thermoelectric transport under the strong electron-phonon scattering.
\end{abstract}

\maketitle

\section{Introduction}

Thermoelectric conversion, which enables waste heat recovery, is a key technology for resolving the energy crisis.
Enhancing thermoelectric conversion efficiency is a crucial task in this field; accordingly, many studies have been conducted on this topic.
To date, several types of desirable electronic band structures have been proposed, for example, band convergence~\cite{band_conv1, band_conv2}, low-dimensional band dispersion~\cite{low_dim1, low_dim2, low_dim3}, resonant states~\cite{resonant}, and pudding-mold-shaped band structures~\cite{pudding}.
An important feature of these band structures is large density of states and/or large group velocity near the band edge. 
These factors are certainly favorable for efficient thermoelectric conversion when simplification of the scattering process and its strength, e.g., with the constant relaxation-time approximation (CRTA), is validated.

However, scattering can drastically change a situation.
In fact, there are several strategies for enhancing thermoelectric conversion efficiency utilizing scattering, e.g., energy filtering~\cite{energy_filt1, energy_filt2, energy_filt3}, modulation doping~\cite{modulation_dope1, modulation_dope2},
and ionization-impurity scattering~\cite{ion_imp_sc1, ion_imp_sc2}.
Strong electron correlation effects can invoke non-trivial scattering effects, which cause anomalous temperature dependence of the Seebeck coefficient~\cite{cuprate_expt, cuprate_vh1, cuprate_vh2, cuprate_3, cuprate_4}, enhancement of the Seebeck effect by spin entropy~\cite{Koshibae, spin_entropy1, spin_entropy2}, spin fluctuation~\cite{spin_fluc}, (para)magnon drag~\cite{magnon_drag1, magnon_drag2, paramagnon_drag}, scattering by magnetic ions~\cite{mag_scat}, and band renormalization~\cite{FeSb2}.
Recent theoretical developments allow the first-principles treatment of the electron-phonon coupling in transport calculations~\cite{EPW1, EPW2, EPW3, perturbo, elphbolt}.
Using this technique, researchers can investigate, e.g., how intervalley and intravalley electron-phonon scattering differ and affect transport properties~\cite{mobility_polar, interval, Mori_ZrX2, inter_intra_PbXHH, inter_intra_elemental_monolayer}.
It was pointed out that band convergence occurring at distant ${\bm k}$-points is beneficial while that for a single ${\bm k}$-point is not~\cite{when_band_conv}, contrarily to the previous understanding that band convergence is always beneficial assuming simplified scattering processes.
The detrimental effect of band convergence was demonstrated in some materials, e.g., for GaN~\cite{GaN_crystal_field, GaN_crystal_field2}.
Valley engineering to avoid valley degeneracy via strain has also been proposed~\cite{strain_valley}.
The mobility of electrons for characteristic electronic and phonon states in (quasi-)two-dimensional materials have been investigated~\cite{sym_q2d, flexural_phonon, Sb_high_mobility, why_2d_low_mobility}.
The role of interband electron-phonon scattering in electron lifetimes has been discussed in link with highly photoexcited electrons and transport~\cite{hot1, hot2, hot3}.
Enriched knowledge of electron-phonon scattering also leads ones to a strategy for decreasing thermal conductivity via phonon softening that does not degrade electron mobility~\cite{soft_PbTe}.
It is also interesting that the electron-phonon drag enhancement of transport properties has now been analyzed in a first-principles manner~\cite{electron_phonon_drag}.

As a new aspect of electron-phonon scattering, Fedorova {\it et al}.~pointed out that strong interband scattering can invoke an anomalous sign change in the Seebeck coefficient by blurring a portion of the electronic band structure~\cite{ano_el_hole}.
This idea can be used to effectively hide the upper side of the Dirac cone overlapping with a heavy band, which increases the power factor owing to the sharp dispersion of the Dirac cone liberated from the bipolar effect~\cite{dirac_filter}.
However, currently, little is known about such an intriguing role of electron-phonon scattering owing to the theoretical complexities of addressing the very large degrees of freedom in electron-phonon-coupled systems.
In particular, many energy scales appearing there makes it difficult to explore a wide parameter space to find unprecedented phenomena.

\begin{figure*}
\begin{center}
\includegraphics[width=14 cm]{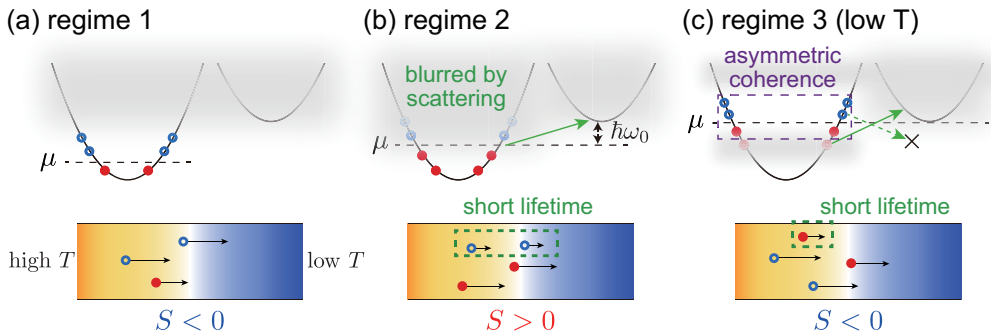}
\caption{A schematic picture showing the three electron transport regimes in the temperature gradient for a two-valley system with strong intervalley scattering.}
\label{fig:regimes}
\end{center}
\end{figure*}

In this paper we analyze a minimal model for a two-electron-valley system with intravalley and intervalley electron-phonon scattering; accordingly, find three electron transport regimes under strong intervalley scattering.
As shown in the schematic represented in Fig.~\ref{fig:regimes}, the usual electron transport with a negative Seebeck coefficient $S<0$ is realized in {\it regime 1}, where the chemical potential $\mu$ is placed near the band edge and far from the other valley.
In this regime, the Seebeck effect is dominated by carriers above the chemical potential (electron carriers) owing to their larger concentration and group velocity than those of the carriers below the chemical potential (hole carriers), which yields $S<0$. In {\it regime 2}, $S>0$ is realized by strong intervalley scattering significantly shortening the lifetime of electron carriers~\cite{ano_el_hole}, while hole carriers are energetically far from the other band edge so that they do not suffer from intervalley scattering.
In addition, we find that reentrant to $S<0$ and enhancement of PF, called {\it regime 3}, is realized when $\mu$ is placed near the edge of the other electron valley at low temperature. The PF enhancement of this regime is caused by {\it asymmetric coherence} where only hole carriers suffer from intervalley scattering effects.
Our finding shed light on an unexplored role of the electron-phonon coupling, and will trigger a search for high-performance thermoelectric materials from a new perspective.

\section{Methods}
We used a two-dimensional effective model of an electron-phonon coupled system expressed as follows:
\begin{align}
\mathcal{H} = &\sum_{\bm{k}=(k_x,k_y)} \sum_{\sigma \in \{\uparrow, \downarrow\}} \sum_{i=1}^2 \epsilon_{{\bm k} i} \hat{c}^{\dag}_{\bm{k}\sigma i} \hat{c}_{\bm{k}\sigma i}\notag \\
&+\sum_{\bm{q}=(q_x,q_y)} \sum_{\nu=1}^2 \hbar \omega_{{\bm q} \nu} \left( \hat{b}^{\dag}_{\bm{q}\nu} \hat{b}_{\bm{q}\nu} + \frac{1}{2} \right)\notag \\
&+\frac{1}{\sqrt{N}}\sum_{\bm{k},\bm{q},\sigma, i_1, i_2, \nu}\bigg[ g_{{\bm q} i_1 i_2 \nu} \hat{c}^{\dag}_{{\bm k}+{\bm q} \sigma i_1} \hat{c}_{\bm{k} \sigma i_2}\hat{b}_{\bm{q} \nu} + h.c. \bigg],
\end{align}
where $\sigma \in \{\uparrow, \downarrow\}$ is the spin index, $N$ is the number of ${\bm k}$ (${\bm q}$)-points in our simulation, $\hat{c}$ ($\hat{c}^{\dag}$) and $\hat{b}$ ($\hat{b}^{\dag}$) are the annihilation (creation) operators of an electron and a phonon, respectively. The wave numbers of electrons and phonons satisfy $-\pi/a \leq k_x, k_y, q_x, q_y \leq \pi/a$, where $a$ is the lattice constant.
The electron band dispersion was given as
\begin{align}
&\epsilon_{{\bm k}i} = \notag\\
&\begin{cases}
-\frac{\hbar^2}{m^* a^2} (\cos (k_x a) + \cos (k_y a) - 2) &(i=1)\\
-\frac{\hbar^2}{m^* a^2} (\cos (k_x a+\pi) + \cos (k_y a+\pi) - 2) + \Delta &(i=2)
\end{cases}\label{eq:electron_band}
\end{align}
where $\Delta$ and $m^*$ are the energy offset between the two valleys and the effective mass, respectively.
The phonon band dispersion was given as
\begin{equation}
\omega_{{\bm q} \nu} =
\begin{cases}
v_0 \sqrt{q_x^2+q_y^2} & (\nu = 1)\\
\omega_0 & (\nu = 2)
\end{cases}
\end{equation}
where $v_0$ and $\omega_0$ are the acoustic phonon velocity and Einstein phonon frequency, respectively.
A simple electron-phonon coupling was assumed as follows:
\begin{equation}
g_{{\bm q} i_1 i_2 \nu} = g_{\mathrm{A}} \sqrt{|{\bm q}|a} \delta_{i_1 i_2}\delta_{\nu 1} + g_{\mathrm{E}} (1-\delta_{i_1 i_2}) \delta_{\nu 2}, \label{eq:g}
\end{equation}
where the coupling constants for intravalley scattering ($i_1=i_2$) by the acoustic phonon ($\nu=1$) and intervalley scattering ($i_1\neq i_2$) by the Einstein phonon ($\nu=2$) are $g_{\mathrm{A}}$ and $g_{\mathrm{E}}$, respectively.
This is a minimal model representing a two-electron-valley system with electron-phonon coupling.
The $\sqrt{|{\bm q}|}$ dependence of the acoustic phonon expressed in Eq.~(\ref{eq:g}) was assumed by considering $g\propto M_{{\bm q} {\bm k}} \omega_{{\bm q}, \nu=1}^{-1/2} \propto M_{{\bm q} {\bm k}} |{\bm q}|^{-1/2}$ with the matrix element $M_{{\bm q} {\bm k}}$ for the potential variation $\delta V_{\bm q}$ associated with the phonon mode satisfying $M_{{\bm q} {\bm k}} = \langle {\bm k}+{\bm q}|\delta V_{\bm q}|{\bm k}\rangle \propto |{\bm q}|$, which holds, e.g., for the deformation potential approximation~\cite{Mahan}.
Note that electron-phonon coupling for longitudinal optical phonons becomes very strong around $|{\bm q}|=0$ for polar materials, as is well known as Fr{\"o}hlich coupling.
In our model, two electron valleys are placed at different ${\bm k}$-points and the chemical potential is far from band crossing, that is, momentum transfer $|{\bm q}|$ for the intervalley scattering is sufficiently large so that we can neglect $q$-dependence of the electron-phonon coupling for $\nu=2$.

Transport calculations were performed based on the Boltzmann transport theory.
The transport coefficient $K_j$ ($j=0, 1$) is defined as follows:
\begin{equation}
K_j = -\frac{2}{\Omega N}\sum_{{\bm k},i} \tau_{{\bm k} i} v_{x;{\bm k} i}^2 (\epsilon_{{\bm k} i}- \mu)^j  \frac{\partial f_{{\bm k}i}}{\partial \epsilon}, \label{eq:Kdef}
\end{equation}
where $\mu$ is the chemical potential, $f_{{\bm k}i}=  (e^{\beta (\epsilon_{{\bm k}i} - \mu)} + 1)^{-1}$ is the Fermi-Dirac distribution function for the inverse temperature $\beta = (k_B T)^{-1}$, $v_{x;{\bm k} i}$ is the $x$-component of the group velocity ${\bm v}_{{\bm k} i} = \hbar^{-1}\partial \epsilon_{{\bm k} i}/\partial {\bm k}$, $\Omega=a^3$ is a unit-cell volume, and factor of two on the right-hand side comes from spin degeneracy.
Here, we used the momentum-relaxation time approximation, and then the electron relaxation time $\tau_{{\bm k},i}$ was calculated using the following equation~\cite{tau_eq1, tau_eq2},
\begin{align}
\frac{1}{\tau_{{\bm k} i}} = \frac{2\pi}{\hbar N} \sum_{{\bm q}, i',\nu} \left( 1 - \frac{{\bm v}_{{\bm k} i} \cdot {\bm v}_{{\bm k}+{\bm q} i'} }{|{\bm v}_{{\bm k} i}  | | {\bm v}_{{\bm k}+{\bm q} i'} |}\right) |g_{{\bm q} i i' \nu}|^2 \notag\\
\times \bigg[ W^{(+)}_{{\bm k} {\bm q} i i' \nu} + W^{(-)}_{{\bm k} {\bm q} i i' \nu} \bigg] \label{eq:tau_equation}
\end{align}
with
\begin{equation}
W^{(\pm)}_{{\bm k} {\bm q} i i' \nu} = \delta(\epsilon_{{\bm k}i}- \epsilon_{{\bm k}+{\bm q} i'} \pm \hbar \omega_{{\bm q} \nu}) 
\begin{cases}
f_{{\bm k}+{\bm q} i'} + n_{{\bm q} \nu}\\
1 - f_{{\bm k}+{\bm q} i'} + n_{{\bm q} \nu}
\end{cases},\label{eq:Wdef}
\end{equation}
where $n_{{\bm q}\nu} = (e^{\beta \hbar \omega_{{\bm q}\nu}} - 1)^{-1}$ is the Bose-Einstein distribution function.
The electrical conductivity $\sigma$, Seebeck coefficient $S$, and power factor PF were calculated as follows:
\begin{equation}
\sigma = e^2 K_0,\ \ S = -\frac{1}{eT}\frac{K_1}{K_0},\ \ \mathrm{PF} = \sigma S^2. \label{eq:trans}
\end{equation}
The electron-phonon coupling affects the electron transport only through the electron relaxation time in this formulation.
Renormalization effects through the real part of the electron self energy is an important future issue.
Possible effects of scattering processes other than the electron-phonon scattering are discussed in Sec.~\ref{sec:real_materials}.

We used $\epsilon_0 \equiv \hbar^2(m^* a^2)^{-1}$ in the electron band dispersion, Eq.~(\ref{eq:electron_band}), as an energy unit. Then, we fixed $g_{\mathrm{A}}=\epsilon_0$ and used $g_{\mathrm{E}}/g_{\mathrm{A}}$ as a parameter representing the strength of the intervalley scattering.
We used $\hbar\omega_0=0.2 \epsilon_0$ so that the phonon energy is an order of magnitude smaller than the electronic bandwidth. The acoustic phonon velocity was set as $v_0 = \omega_0 a \pi^{-1}$ so that $\omega_{{\bm q} 1} \sim \omega_{{\bm q} 2}$ holds near the Brillouin zone boundary.
We used $500 \times 500$ and $1,000 \times 1,000$ ${\bm k}$ (${\bm q}$)-meshes for $k_B T\geq 0.25 \epsilon_0$ and $k_B T<0.25 \epsilon_0$, respectively, except $\tau$-plots where a $2,400 \times 2,400$ ${\bm k}$ (${\bm q}$)-mesh was used.
The delta function appearing in Eq.~(\ref{eq:Wdef}) was approximated as a Gaussian distribution function with a broadening energy width of $0.001\epsilon_0$.

\section{Results and Discussions}
\subsection{Three regimes for electron transport}

\begin{figure*}
\begin{center}
\includegraphics[width=12 cm]{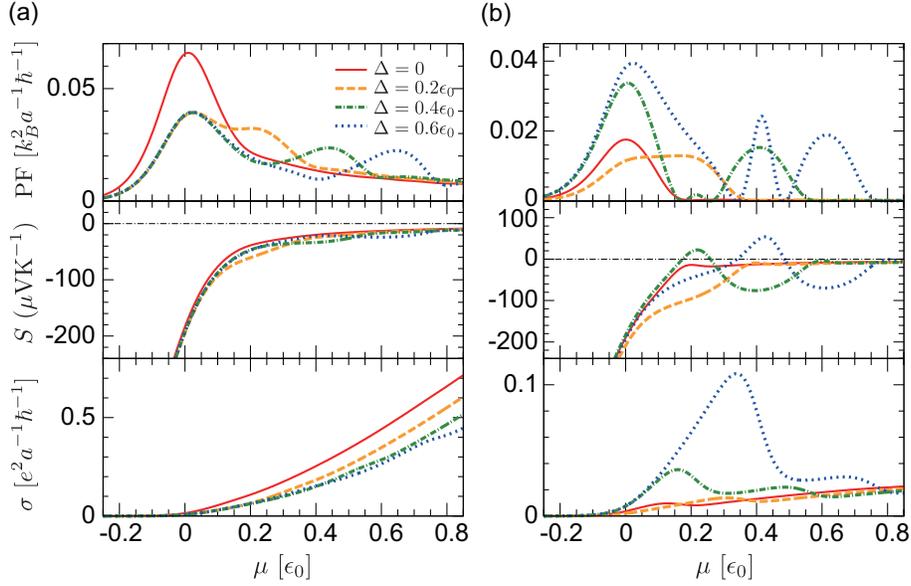}
\caption{Calculated transport properties using $k_B T=0.04 \epsilon_0$ and $g_{\mathrm{E}}/g_{\mathrm{A}}=$ 2 and 20 for panels (a) and (b), respectively.}
\label{fig:2}
\end{center}
\end{figure*}

First, we present the calculated transport properties using $g_{\mathrm{E}}/g_{\mathrm{A}}=2$ and 20.
Figure~\ref{fig:2}(a) presents the calculated PF with various $\Delta$ using $k_BT=0.04 \epsilon_0$ and $g_{\mathrm{E}}/g_{\mathrm{A}}=2$, where the electron-phonon couplings for intervalley and intravalley scattering have a comparable strength.
In this case, a high PF was obtained for $\Delta=0$, i.e., when two valleys are degenerate.
For $\Delta \neq 0$, two PF peaks appear near the band edges of the two valleys, $\mu=0$ and $\Delta$. $S<0$ always holds.
Note that we use a unit of $\mu$VK$^{-1}$ for $S$ by considering that $S$ in Eq.~(\ref{eq:trans}) is a product of $k_B e^{-1} = 86.17\ \mu$VK$^{-1}$ and a dimensionless quantity $-(k_B T)^{-1} K_1 K_0^{-1}$, as is often done in model calculations.
These observations are consistent with many transport calculations using CRTA.

However, as shown in Fig.~\ref{fig:2}(b), where intervalley scattering is much stronger than intravalley scattering, $g_{\mathrm{E}}/g_{\mathrm{A}}=20$, the situation is quite different. First, the band degeneracy at $\Delta =0$ yields the smallest PF peak, which sharply contrasts the case with $g_{\mathrm{E}}/g_{\mathrm{A}}=2$.
This is because valley degeneracy significantly shortens the electron relaxation time via intervalley scattering.
Band convergence, $\Delta = 0$, is no longer a good strategy for enhancing PF under strong intervalley scattering (see, e.g., Ref.~\onlinecite{strain_valley} for mobility degradation via band convergence).
In addition, PF exhibits a remarkable three-peaked structure for large $\Delta$, such as $\Delta = 0.6\epsilon_0$ in Fig.~\ref{fig:2}(b).
These three PF peaks appear at approximately $\mu=0$, $\Delta - \hbar \omega_0\ (=0.4\epsilon_0\ \mathrm{for}\ \Delta = 0.6\epsilon_0)$, and $\Delta$.
Around the second peak, the Seebeck coefficient exhibits a characteristic sign change, as reported in Ref.~\onlinecite{ano_el_hole}.
Hereafter, we denote the transport regimes around these three $\mu$ values as {\it regimes 1, 2, 3}.

\subsection{$\Delta$--$\mu$ plot}

Before interpreting the three-peaked PF structure shown in Fig.~\ref{fig:2}(b) with $\Delta=0.6 \epsilon_0$, we shall answer a natural question that arises here: How robust is the three-peaked structure? In fact, this interesting PF structure strongly depends on temperature.

\begin{figure*}
\begin{center}
\includegraphics[width=14 cm]{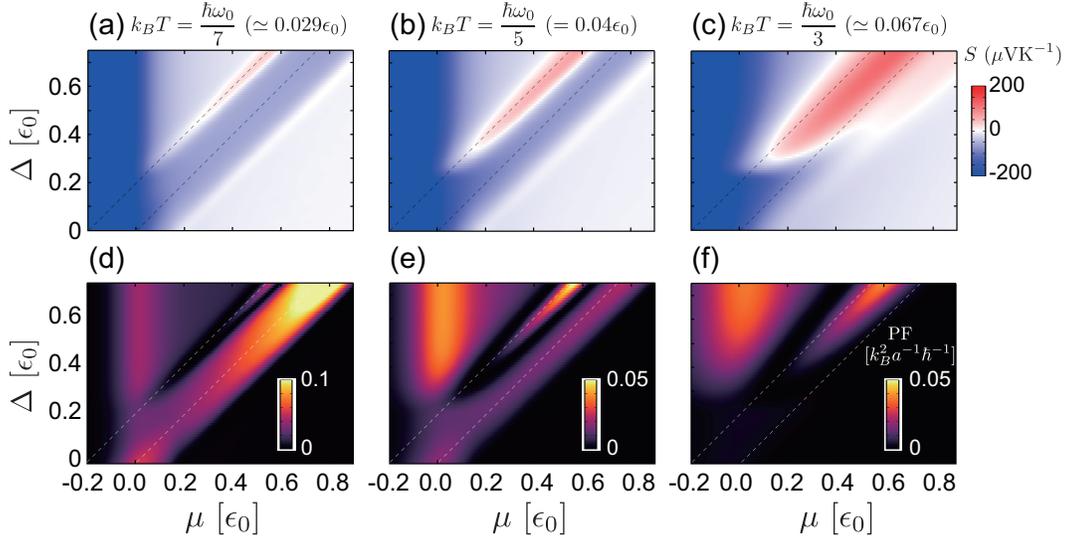}
\caption{$\Delta$--$\mu$ plots for (a)--(c) $S$ and (d)--(f) PF calculated using $g_{\mathrm{E}}/g_{\mathrm{A}}=20$ and three temperatures: (a)(d) $k_B T=\hbar \omega_0/7\ (\simeq 0.029 \epsilon_0)$, (b)(e) $k_B T= \hbar \omega_0/5\ (=0.04 \epsilon_0)$, and (c)(f) $k_B T= \hbar \omega_0/3\ (\simeq 0.067 \epsilon_0 )$. Broken lines $\mu = \Delta$ and $\mu = \Delta - \hbar\omega_0\ ( = \Delta - 0.2 \epsilon_0)$ are shown as guides to the eyes.}
\label{fig:3}
\end{center}
\end{figure*}

Figure~\ref{fig:3} presents $S$ and PF values calculated using $g_{\mathrm{E}}/g_{\mathrm{A}}=20$ and three temperatures: $k_B T= \hbar \omega_0/7\ (\simeq 0.029 \epsilon_0)$, $\hbar\omega_0/5\ (=0.04 \epsilon_0)$, and $\hbar \omega_0/3\ (\simeq 0.067\epsilon_0)$. In these plots, we varied both the chemical potential $\mu$ and the electron-valley offset $\Delta$.
The PF peak in {\it regime 1} around $\mu\sim 0$ is relatively robust while the peak value itself can be small for a small $\Delta$.
On the other hand, the PF peak in {\it regime 2} around $\mu\sim \Delta - \hbar \omega_0$ is conspicuous at high $T$ but diminishes by lowering $T$.
Note that {\it regime} 2 is identified by $S>0$ regions in Figs.~\ref{fig:3}(a)--(c).
The PF peak in {\it regime 3} around $\mu\sim \Delta$ shows an opposite trend: it does not appear at high $T$, e.g., in Fig.~\ref{fig:3}(f), but develops by lowering $T$, which finally offers higher PF values than PF peak values in {\it regimes 1} and {\it 2} at $k_B T= \hbar \omega_0/7$, as shown in Fig.~\ref{fig:3}(d).

\subsection{Electron relaxation time~\label{sec:tau}}

\begin{figure*}
\begin{center}
\includegraphics[width=14 cm]{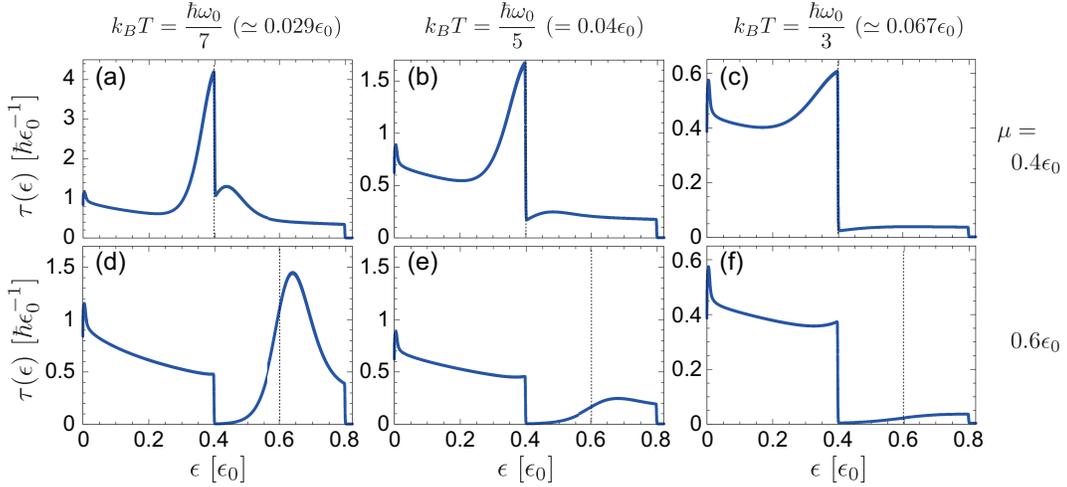} 
\caption{Electron relaxation time $\tau(\epsilon)$ for the electron valley $i=1$ calculated using $\Delta=0.6 \epsilon_0$ and $g_{\mathrm{E}}/g_{\mathrm{A}}=20$. Chemical potential was set as (a)--(c) $\mu=\Delta - \hbar \omega_0\ (=0.4 \epsilon_0)$ and (d)--(f) $\mu=\Delta\ (=0.6 \epsilon_0)$. Temperature was set as  (a)(d) $k_B T= \hbar \omega_0/7\ (\simeq 0.029 \epsilon_0)$, (b)(e) $k_B T= \hbar \omega_0/5\ (=0.04 \epsilon_0)$, and (c)(f) $k_B T= \hbar \omega_0/3\ (\simeq 0.067 \epsilon_0)$. Broken lines denote the position of the chemical potential, $\epsilon = \mu$.}
\label{fig:4}
\end{center}
\end{figure*}

To understand the mechanism of how the three-peaked structure of PF occurs, we calculated the electron relaxation time $\tau_{{\bm k}i}$ as a function of the corresponding electron energy $\epsilon_{{\bm k}i}$: $\tau(\epsilon)$, as shown in Fig.~\ref{fig:4}.
Here, we only show the electron relaxation time of the $i=1$ valley because the electrons in valley 2 contributes little to the transport coefficients $K_j$ ($j=1,2$) for the chemical potentials used here [see, Sec.~\ref{sec:discussion_regime3}].
The calculation was performed using $\Delta=0.6 \epsilon_0$, $g_{\mathrm{E}}/g_{\mathrm{A}}=20$, $\mu=\Delta - \hbar\omega_0\ (=0.4 \epsilon_0)$ and $\Delta\ (=0.6 \epsilon_0)$, and $k_B T= \hbar \omega_0/7\ (\simeq 0.029 \epsilon_0)$, $\hbar \omega_0/5\ (=0.04 \epsilon_0)$, and $\hbar \omega_0/3\ (\simeq 0.067\epsilon_0)$.

At high temperatures, $k_B T \gg \hbar \omega_0$, Eq.~(\ref{eq:Wdef}) can be approximately simplified to 
$W^{(\pm)}_{{\bm k} {\bm q} 12 2} \simeq \delta(\epsilon_{{\bm k}1}- \epsilon_{{\bm k}+{\bm q} 2} \pm \hbar\omega_0) n_{{\bm q} 2}$. 
Considering that the $i=2$ valley has the band edge at $\Delta$, i.e., $\epsilon_{{\bm k}+{\bm q} 2}\geq \Delta$, the electron relaxation time $\tau(\epsilon_{{\bm k}1})$ can be approximated as a $\mu$-independent step function: long $\tau$ for $\epsilon_{{\bm k}1} <\Delta - \hbar \omega_0$ where $W^{(\pm)}_{{\bm k} {\bm q} 12 2}\simeq 0$, short $\tau$ for $\Delta - \hbar\omega_0 < \epsilon_{{\bm k}1} <\Delta + \hbar \omega_0$ where $W^{(-)}_{{\bm k} {\bm q} 12 2}\simeq 0$ but $W^{(+)}_{{\bm k} {\bm q} 12 2}$ is activated, and much shorter $\tau$ for $\Delta + \hbar \omega_0 < \epsilon_{{\bm k}1}$ where both $W^{(\pm)}_{{\bm k} {\bm q} 12 2}$ are activated.
In Figs.~\ref{fig:4}(c)(f), while $k_B T = \hbar \omega_0/3$ is not very high, $\tau(\epsilon)$ resembles this step function.
This is why a large $S>0$ was obtained in {\it regime 2}: $\tau$ of electron and hole carriers are sizably different in Fig.~\ref{fig:4}(c).
This situation is illustrated schematically in Fig.~\ref{fig:regimes}(b): only the electron carriers suffer from the strong intervalley scattering~\cite{ano_el_hole}.
This is also similar to the idea of the energy filtering using energy-dependent scattering time~\cite{energy_filt1, energy_filt2}.

At low temperatures, a peak structure of $\tau(\epsilon)$ around $\epsilon\sim \mu$ gradually develops, as shown in Fig.~\ref{fig:4}.
A long-lived (coherent) electron at $|\epsilon - \mu| <  \hbar \omega$ where $\hbar \omega$ is the characteristic phonon energy, is a well-known consequence of the electron-phonon coupling at low temperatures. In fact, considering $n_{{\bm q} \nu}\sim 0$, Eq.~(\ref{eq:Wdef}) becomes $W^{(+)}_{{\bm k} {\bm q} 1i' \nu} \simeq \delta(\epsilon_{{\bm k}1}- \epsilon_{{\bm k}+{\bm q} i'} + \hbar \omega) f_{{\bm k}+{\bm q} i'}$ and $W^{(-)}_{{\bm k} {\bm q} 1i' \nu} \simeq \delta(\epsilon_{{\bm k}1}- \epsilon_{{\bm k}+{\bm q} i'} - \hbar \omega) (1 - f_{{\bm k}+{\bm q} i'} )$, both of which are small for $|\epsilon_{{\bm k}1} - \mu| < \hbar \omega$.
For example, $f_{{\bm k}+{\bm q} i'}$ in $W^{(+)}$ becomes large for $\epsilon_{{\bm k}+{\bm q} i'}<\mu$ and then the $\delta$-function requires 
$\epsilon_{{\bm k}1} = \epsilon_{{\bm k}+{\bm q} i'} - \hbar \omega < \mu - \hbar\omega$.
For the same reason, $\epsilon_{{\bm k}1} > \mu + \hbar\omega$ is desirable for activating $W^{(-)}$.
Because the temperature broadening of the Fermi-Dirac distribution obscures this tendency, this structure is conspicuous at low temperature.
Note that $\tau(\epsilon)$ ($\epsilon \sim \mu$) at the low-temperature limit has a peak structure due to acoustic-phonon ($\nu=1$) intravalley scattering because acoustic phonons can have an energy smaller than $\hbar \omega_0$.

However, the peak structure of $\tau(\epsilon)$ is remarkably asymmetric around $\epsilon=\mu$ in Fig.~\ref{fig:4}(d) for the following reason.
Hole carriers with $\Delta - \hbar \omega_0 < \epsilon_{{\bm k}1}<\mu$ suffer from scattering by $W^{(+)}_{{\bm k} {\bm q} 12 2} \simeq \delta(\epsilon_{{\bm k}1}- \epsilon_{{\bm k}+{\bm q} 2} + \hbar \omega_0) f_{{\bm k}+{\bm q} 2}$ owing to the small but non-zero $f_{{\bm k}+{\bm q} 2}$ for unoccupied states with $\epsilon_{{\bm k}+{\bm q} 2}=\epsilon_{{\bm k}1}+ \hbar \omega_0 > \Delta$. Electron carriers also suffer from this scattering but the effect is weaker because of the smaller $f_{{\bm k}+{\bm q} 2}$.
On the other hand, $W^{(-)}_{{\bm k} {\bm q} 12 2} \simeq \delta(\epsilon_{{\bm k}1}- \epsilon_{{\bm k}+{\bm q} 2} - \hbar\omega_0) (1 - f_{{\bm k}+{\bm q} 2} )$ is prohibited for both hole and electron carriers with $\epsilon_{{\bm k}1}<\mu + \hbar\omega_0$, because the $\delta$-function requires $\epsilon_{{\bm k}1} = \epsilon_{{\bm k}+{\bm q} 2} + \hbar \omega_0 \geq \Delta + \hbar \omega_0 = \mu + \hbar \omega_0$.
Therefore, electron carriers have a longer relaxation time than that for hole carriers, which yields a large $|S|$ with a negative sign.
This {\it asymmetric coherence} is the origin of the {\it regime 3}, as schematically shown in Fig.~\ref{fig:regimes}(c).

\subsection{Temperature dependence}

\begin{figure}
\begin{center}
\includegraphics[width=8.3 cm]{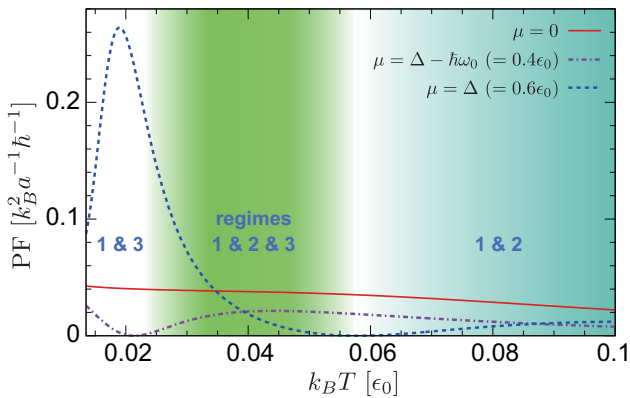}
\caption{Temperature dependence of PF calculated using $\Delta=0.6 \epsilon_0$ and $g_{\mathrm{E}}/g_{\mathrm{A}}=20$.}
\label{fig:5}
\end{center}
\end{figure}

We point out that the three-peaked structure of PF exhibits a characteristic temperature dependence.
Figure~\ref{fig:5} presents the temperature dependence of PF calculated using $\Delta=0.6 \epsilon_0$ and $g_{\mathrm{E}}/g_{\mathrm{A}}=20$.
For {\it regime 2}, PF at $\mu=\Delta - \hbar \omega_0$ becomes zero at $k_B T\sim 0.02 \epsilon_0$, under which the Seebeck coefficient becomes negative and {\it regime 2} disappears. This is because the coherent peak of $\tau$ develops by lowering the temperature, which conceals the step-like structure of $\tau(\epsilon)$ as seen in Figs.~\ref{fig:4}(a)--(c).
For {\it regime 3}, PF at $\mu=\Delta$ becomes zero at $k_B T\sim 0.06 \epsilon_0$, above which the Seebeck coefficient becomes positive and {\it regime 3} is absorbed in {\it regime 2}. As discussed in the previous paragraph, the origin of {\it regime 3}, namely, asymmetric coherence of the electron relaxation time, does not occur at high temperatures.
It is also remarkable that PF peak in {\it regime 3} is very large at $k_B T\sim 0.02 \epsilon_0$. Since this PF value is much larger than that in {\it regime 1}, we can say that PF is {\it enhanced} by strong intervalley scattering here, contrary to the common understanding that the scattering has a detrimental effect on transport.
However, we note that scattering processes not considered here, such as the impurity scattering, possibly become dominant in low temperatures, which will significantly suppress such a strong enhancement of PF in real materials. A link to real materials shall be discussed in Sec.~\ref{sec:real_materials}.

\subsection{Discussions}

\subsubsection{Origin of PF enhancement near the second-valley bottom $(\mu = \Delta)$~\label{sec:discussion_regime3}}

One of interesting phenomena caused by strong intervalley scattering is the Gunn effect, where electron carriers that are originally populated in one valley become able to reach other higher-energy valleys by applying the strong electric field.
From the viewpoint that carriers in each valley can participate in transport, it is important to clarify that carriers in which valley dominate electron transport in {\it regime 3} where the chemical potential reaches the bottom of the second valley.
To investigate this point, we defined the following transport coefficient,
\begin{equation}
\tilde{K}_j = -\frac{2}{\Omega N} \sum_{{\bm k}} \tau_{{\bm k}1} v_{x;{\bm k} 1}^2 (\epsilon_{{\bm k} 1}- \mu)^j  \frac{\partial f_{{\bm k}1}}{\partial \epsilon}, \label{eq:K_tilde_def}
\end{equation}
where the summation over the valley index $i$ in Eq.~(\ref{eq:Kdef}) is restricted to $i=1$. In other words, only carriers in the $i=1$ valley participate in transport for Eq.~(\ref{eq:K_tilde_def}). By using $\tilde{K}_j$ (Eq.~(\ref{eq:K_tilde_def})) instead of $K_j$ (Eq.~(\ref{eq:Kdef})), we can evaluate the contribution of the $i=1$ valley to transport.

\begin{figure*}
\begin{center}
\includegraphics[width=12 cm]{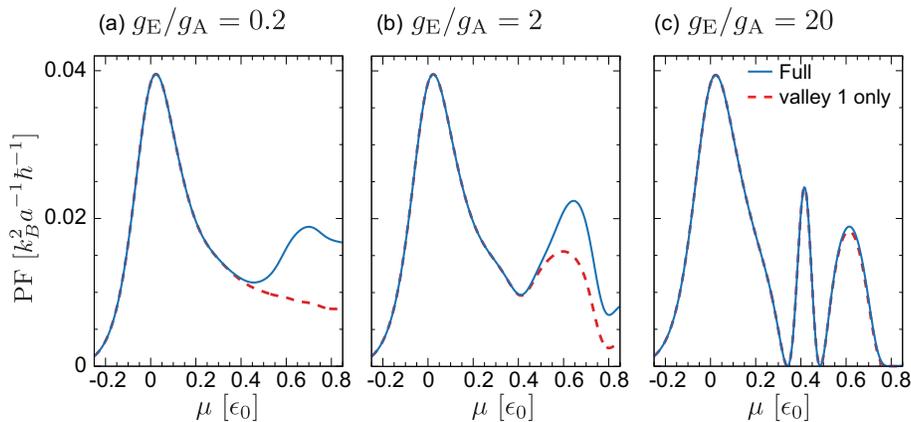}
\caption{PF plots using $\Delta=0.6\epsilon_0$, $k_B T=0.04 \epsilon_0$, and $g_{\mathrm{E}}/g_{\mathrm{A}}=$ (a) 0.2, (b) 2, and (c) 20. Blue solid and red broken lines show full calculation results and those using $\tilde{K}$ (Eq.~(\ref{eq:K_tilde_def})) where only carriers in the $i=1$ valley  participate in transport, respectively.}
\label{fig:6}
\end{center}
\end{figure*}

In Fig.~\ref{fig:6}, PF values calculated using $\Delta = 0.6 \epsilon_0$, $k_B T=0.04 \epsilon_0$, and $g_{\mathrm{E}}/g_{\mathrm{A}} = 0.2$, $2$, and $20$, are plotted against the chemical potential $\mu$. Red broken lines denoted as ``valley 1 only'' represent PF calculated using $\tilde{K}_j$ instead of $K_j$. As shown in Fig.~\ref{fig:6}(a), under weak intervalley scattering, the PF peak at $\mu \sim \Delta\ (=0.6 \epsilon_0)$ forms by the second-valley contribution. This is verified by the fact that the ``valley 1 only'' line does not have a peak at around $\mu=\Delta$. On the other hand, under strong intervalley scattering, the PF peak at $\mu \sim \Delta$ solely originates from the first-valley carriers as shown in Fig.~\ref{fig:6}(c). To say, the $i=2$ valley just acts as a scatterer there. While the PF peaks at $\mu\sim \Delta$ similarly appear regardless the strength of the intervalley scattering in Figs.~\ref{fig:6}(a)--(c), their origins are different in weak and strong intervalley scattering regimes; Carriers in the second valley enhances PF under weak intervalley scattering while the strong electron-hole asymmetry of electron relaxation time in the first valley results in the PF peak under strong intervalley scattering.

\subsubsection{Breakdown of CRTA in regimes 2 and 3}

In theoretical studies on electron transport, simple approximations for electron relaxation time such as CRTA are often adopted.
In some studies, $\tau (\epsilon) \propto \epsilon^r$ is adopted (e.g., Ref.~\onlinecite{tau_approx}) where $r$ is an exponent.
In the following, we discuss when the simple approximation for electron relaxation time breaks down in our model, because strong energy dependence of $\tau(\epsilon)$ is crucial for PF enhancement investigated in our study. We consider the case of $r=0$, i.e., CRTA, because  $\tau^{-1}$ is roughly proportional to density of states (DOS) and DOS is roughly constant near the band edge in two-dimensional systems.

\begin{figure*}
\begin{center}
\includegraphics[width=18 cm]{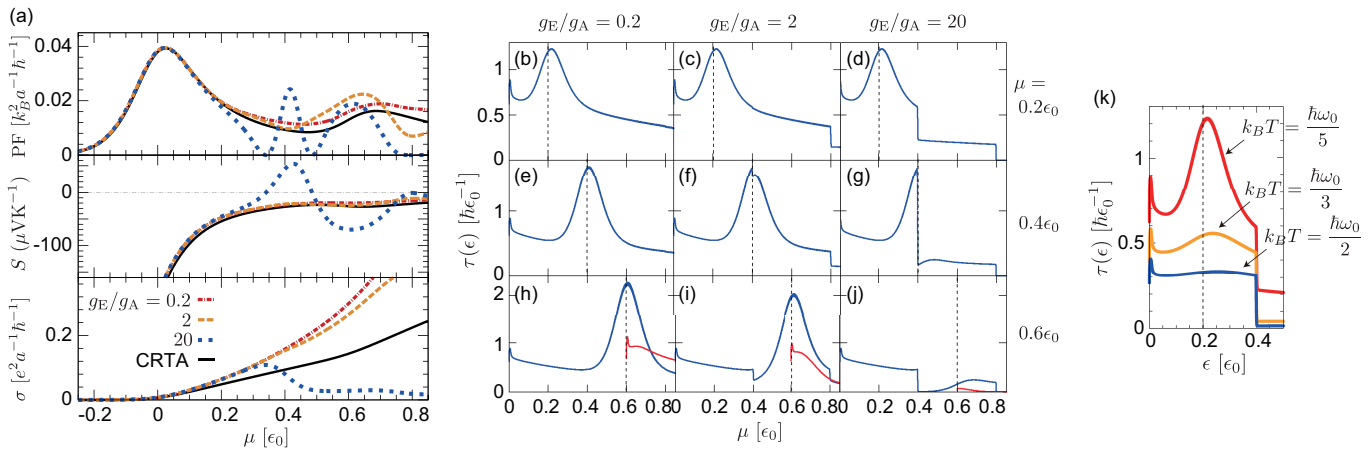}
\caption{a) Calculated transport properties using $k_B T=0.04 \epsilon_0$ and $\Delta = 0.6 \epsilon_0$. The red, orange, blue, and black lines represent calculation data using $g_{\mathrm{E}}/g_{\mathrm{A}}=0.2$, $2$, $20$, and CRTA with $\tau=0.78 \hbar^{-1}\epsilon_0$, respectively.
(b)--(j) Electron relaxation time $\tau(\epsilon)$ calculated using $k_B T=0.04 \epsilon_0$ and $\Delta = 0.6 \epsilon_0$.
Chemical potential was set as (b)--(d) $\mu=0.2 \epsilon_0$, (e)--(g) $\mu=0.4 \epsilon_0$, (h)--(j) $\mu=0.6 \epsilon_0$, respectively.
$g_{\mathrm{E}}/g_{\mathrm{A}}$ was set as (b)(e)(h) $0.2$, (c)(f)(i) $2$, and (d)(g)(j) $20$, respectively.
While $\tau(\epsilon)$ only for the first ($i=1$) valley is shown in panels (b)--(g), $\tau(\epsilon)$ for the first ($i=1$) and the second ($i=2$) valleys are shown with blue and red color, respectively, in panels (h)--(j).
(k) $\tau(\epsilon)$ calculated using $\Delta = 0.6 \epsilon_0$ and $\mu=0.2 \epsilon_0$. The red, orange, and blue lines represent calculation data using $k_B T= \hbar \omega_0/5\ (=0.04 \epsilon_0)$, $\hbar \omega_0/3\ (\simeq 0.067 \epsilon_0)$, and $\hbar \omega_0/2\ (=0.1 \epsilon_0)$, respectively.}
\label{fig:7}
\end{center}
\end{figure*}

First, we present calculated transport properties using $k_B T= 0.04 \epsilon_0$ and $\Delta = 0.6 \epsilon_0$ in Fig.~\ref{fig:7}(a).
For small $\mu$ ($\mu<0.2\epsilon_0$), all calculation results, $g_{\mathrm{E}}/g_{\mathrm{A}}=0.2, 2, 20$ and CRTA, agree well, which means that CRTA is valid near the band edge.
Here, transport for such $\mu$ that is far away from the second-band bottom, $\Delta$, is almost in the single-band regime and so is almost independent of $g_{\mathrm{E}}$.
Thus, $\tau$ for CRTA was determined so that the PF peak height near $\mu=0$ is consistent with our calculation results using several $g_{\mathrm{E}}/g_{\mathrm{A}}$ shown here: $\tau=0.78 \hbar^{-1}\epsilon_0$.
On the other hand, for larger $\mu$, the characteristic PF peaks and a sign change of the Seebeck coefficient are found for $g_{\mathrm{E}}/g_{\mathrm{A}}=20$ as we have seen in Fig.~\ref{fig:2}(b), while these features are absent in calculated data using $g_{\mathrm{E}}/g_{\mathrm{A}}=0.2, 2$, and CRTA.
This is natural considering that the origin of the PF enhancement in {\it regimes 2} and {\it 3} is a sharp change of $\tau(\epsilon)$ near the chemical potential as we have discussed in Sec.~\ref{sec:tau}.
Thus, CRTA cannot describe the PF enhancement in {\it regimes 2} and {\it 3} while CRTA is valid for {\it regime 1} or for the cases where $g_{\mathrm{E}}/g_{\mathrm{A}}$ is not so large. We note that CRTA using different $\tau$ for the two valleys does not change the conclusion here; changing $\tau$ of the second valley just changes the height of the PF peak near $\mu=\Delta$. We also note that the PF peak near $\mu=\Delta$ for CRTA or small $g_{\mathrm{E}}/g_{\mathrm{A}}$ has a different origin from that for $g_{\mathrm{E}}/g_{\mathrm{A}}=20$ as discussed in Sec.~\ref{sec:discussion_regime3}.

Next, we directly compare the calculated relaxation time using $k_B T= 0.04 \epsilon_0$ and $\Delta = 0.6 \epsilon_0$ in Figs.~\ref{fig:7}(b)--(j).
For $\mu = 0.2 \epsilon_0$ shown in Figs.~\ref{fig:7}(b)--(d), intervalley scattering is almost absent near the chemical potential because the band edge of the second valley, $\Delta$, is much higher than the chemical potential. Therefore, $\tau(\epsilon)$ near the chemical potential $\mu = 0.2 \epsilon_0$ has a similar shape among different $g_{\mathrm{E}}/g_{\mathrm{A}}$.
We note that $\tau(\epsilon)$ is not so energy-independent as assumed in CRTA, even in this regime. A peaked structure of $\tau(\epsilon)$ is weakened at high temperatures as shown in Fig.~\ref{fig:7}(k), which compares $\tau(\epsilon)$ calculated using $\Delta = 0.6 \epsilon_0$ and $\mu=0.2 \epsilon_0$ among three temperatures.
Thus, CRTA is better validated in high temperature, while PF is rather insensitive to an energy dependence of $\tau(\epsilon)$ even in lower temperatures.
For $\mu = 0.4 \epsilon_0$ shown in Figs.~\ref{fig:7}(e)--(g), a rapid drop of $\tau(\epsilon)$ near the chemical potential is prominent for $g_{\mathrm{E}}/g_{\mathrm{A}}=20$ in Fig.~\ref{fig:7}(g), while it is not so for $g_{\mathrm{E}}/g_{\mathrm{A}}=2$ in Fig.~\ref{fig:7}(f) and almost discernible for $g_{\mathrm{E}}/g_{\mathrm{A}}=0.2$ in Fig.~\ref{fig:7}(e).
As we have seen in Fig.~\ref{fig:7}(a), CRTA becomes invalid for strong $g_{\mathrm{E}}/g_{\mathrm{A}}$ with the chemical potential lying near the second-band bottom.
The situation is similar for $\mu = 0.6 \epsilon_0$ as shown in Fig.~\ref{fig:7}(h)--(j).
While $\tau(\epsilon)$ does not change very rapidly near the chemical potential in Fig.~\ref{fig:7}(j), 
CRTA is still invalid since the electron-hole-asymmetric $\tau(\epsilon)$ is a key to form the PF peak in {\it regime 3} as discussed in Sec.~\ref{sec:discussion_regime3}.

\subsubsection{Link to real materials~\label{sec:real_materials}}

Finally, we discuss how our study,  specifically, PF enhancement in {\it regimes 2} and {\it 3}, can be realized in materials. Below, we list several important requirements.

{\it Sizable band offset $\Delta$.---}
As can be seen in Fig.~\ref{fig:3}, we need a sizable band offset $\Delta$ to get PF enhancement in {\it regimes 2} and {\it 3}.
There are many studies of thermoelectric materials that aim to control an electronic band structure and realize, e.g., a multi-valley band structure.
For example, an energy offset between two different $p$ bands in Zintl compounds can be controlled via atomic replacement (e.g., Ref.~\onlinecite{when_band_conv}).
Thus, it will be possible to control $\Delta$ in real materials.

{\it Strong intervalley scattering.---}
The key ingredient for PF enhancement in {\it regimes 2} and {\it 3} is the strong intervalley scattering. To realize this, a straightforward way is to find out materials with strong intervalley electron-phonon coupling.
We raise two possibilities: (i) materials having soft phonon modes for an intervalley wave vector ${\bm q}$, which tend to have strong electron-phonon coupling, and (ii) materials having two electron valleys at the same ${\bm k}$-point with some energy offset, where the strong Fr{\"o}hlich coupling for ${\bm q}\sim {\bm 0}$ can be used as the intervalley scattering.

Interestingly, materials with moderate intervalley electron-phonon coupling can also be good candidates when the second valley has high DOS owing to, e.g., valley multiplicity, heavy effective mass, and low-dimensionality as pointed out in Ref.~\onlinecite{ano_el_hole}. Because the second valley in our analysis plays a role of the scatterer, PF should simply benefit from high DOS of the second valley. 
It is worth noting that, in recent theoretical calculation~\cite{ano_el_hole}, $n$-type TaFeSb and $p$-type ZrNiSn exhibit an anomalous sign change of the Seebeck coefficient, which is classified as {\it regime 2} here. In the electronic band structure of TaFeSb, high valley multiplicity and heavy effective mass of the second valley play a key role in the sign change of the Seebeck coefficient. It is also an interesting idea to use a DOS peak of the resonant impurity level~\cite{resonant, resonant_impurity_calc} or localized $f$ levels, e.g., in YbAl$_3$~\cite{YbAl3calc}, as a scatterer.

{\it Low temperature.---}
We should consider a low-temperature regime, in particular for {\it regime 3}, which appears below $k_B T (\hbar \omega_0)^{-1} \sim 0.2\ (=0.04/0.2)$ as shown in Fig.~\ref{fig:5}. For example, it amounts to around 100 K for $\hbar \omega_0 = 50$ meV.

In such a low-temperature region, the phono-drag effect can also have an important contribution to the Seebeck coefficient.
While the phonon-drag effect should be small in our minimal model since the optical phonon ($\nu=2$) governing intervalley scattering is dispersionless, 
how to distinguish PF enhancement by our mechanism and that by the phonon-drag effect in general situations is an important future issue.
From this perspective, anomalous transport in {\it regime 2} might be easier to observe since PF enhancement in {\it regime 2} takes place in even higher temperature (see, Fig.~\ref{fig:5}).

{\it Dominant intervalley electron-phonon scattering over other scattering processes.---}
We have assumed that the electron-phonon scattering is dominant over other scattering processes, a validity of which should be carefully examined.
In the following, we discuss two scattering processes that can become strong in heavily doped systems.
In fact, carrier concentration for {\it regimes 2} and {\it 3} amounts to $\sim 10^{21}$ cm$^{-3}$, considering that $\sim 0.1e$ is doped into the unit cell of $\sim 100$ \AA$^3$.

One is the ionized impurity scattering, since ionized impurities are usually introduced to dope carriers into a system.
Ionized impurity scattering is strong when the temperature is low and the impurity concentration is high~\cite{ionized_impurity}.
Another one is the plasmon scattering~\cite{plasmon}, which is strong when carrier concentration is high~\cite{plasmon, plasmon2}, e.g., $> 10^{19}$ cm$^{-3}$ for bulk silicon~\cite{plasmon2}.
Since long-rangeness of the Coulomb interaction is a key for both scattering mechanisms, scattering with a wave vector ${\bm q}\sim {\bm 0}$ is strong.
This feature can enhance the intravalley scattering, which is also an undesirable aspect.

Considering the existence of these scattering channels, realization of our idea in materials does not seem to be easy. 
One possibility is to consider very strong intervalley electron-phonon scattering that overwhelms these scattering processes owing to a very high DOS peak of localized states such as impurity levels or $f$ bands as we have discussed in this section.
Another possible candidate we propose here is an undoped semimetallic system with strong asymmetry between electron and hole pockets in terms of the effective mass (or other features such as the dimensionality and the valley multiplicity).
In this case, PF enhancement will take place near the band edge of the electron or hole pocket with a heavy effective mass; a pocket with the heavy effective mass plays a role of a scatterer as the $i=2$ (second) band in our model, and the other pocket acts as the $i=1$ band in our model.
In semimetallic systems, heavy carrier concentration can be achieved without impurity doping, by which the impurity scattering is suppressed.
In addition, semimetallic state is expected to screen the long-range tail of the Coulomb interaction, which will also suppress both the ionized impurity scattering and the plasmon scattering [see, Refs.~\onlinecite{ionized_impurity, plasmon2}, which show that the coupling strength for these scattering processes becomes week when the dielectric constant of an undoped system becomes large for semiconductors]. 
If carrier concentration is as large as that for metallic systems, the plasmon frequency can become too high, such as $\sim$ 10 eV, to consider as an active scattering channel.
A (semi)metallic state where electron and hole pockets lie at the same ${\bm k}$-point is also a candidate because these scattering processes enhanced at ${\bm q}\sim {\bm 0}$ can be used as a source of the intervalley scattering.
We should carefully check whether these ideas work well in real materials, which is an important and challenging future issue.

\section{Summary}

We have found that electron transport has three regimes under strong intervalley electron-phonon coupling. In addition to the normal transport in {\it regime 1},  significant shortening of $\tau$ above $\Delta - \hbar \omega_0$ and asymmetric coherence by the absence of the scattering paths shown in Fig.~\ref{fig:regimes}(c), invoke {\it regimes 2} and {\it 3}, respectively. A key factor for such PF enhancement is the electron-hole-asymmetric relaxation time realized by strong intervalley scattering.
Our finding gives a clue to find unexplored thermoelectric transport realized by strong electron-phonon coupling.

\acknowledgements
This study was supported by JSPS KAKENHI (Grant Number JP22K04908) and JST FOREST Program (Grant Number JPMJFR212P).

\end{document}